\documentclass[11pt,superscriptaddress,aps,prd,preprint]{revtex4}
\usepackage{amsmath}

\makeatletter
\usepackage[T1]{fontenc}
\usepackage{amsmath}
\usepackage{amssymb}
\usepackage{graphicx}

\usepackage{slashed}

\newcommand{\bea}{\begin{eqnarray}}
\newcommand{\eea}{\end{eqnarray}}

\begin{document}

\title{$f(R)$ gravity and Tsallis holographic dark energy}

\author{P. S. Ens}\email[]{alesandroferreira@fisica.ufmt.br}
\affiliation{Instituto de F\'{\i}sica, Universidade Federal de Mato Grosso,\\
78060-900, Cuiab\'{a}, Mato Grosso, Brazil}

\author{A. F. Santos}\email[]{peter@fisica.ufmt.br}
\affiliation{Instituto de F\'{\i}sica, Universidade Federal de Mato Grosso,\\
78060-900, Cuiab\'{a}, Mato Grosso, Brazil}

\begin{abstract}
 
The $f(R)$ gravity theory is considered. It is a gravitational theory that generalizes the Einstein-Hilbert action. In this context, a holographic dark energy model is studied. Tsallis non-extensive entropy is used to introduce the dark energy density based on the holographic principle. Then the Friedmann equation, deceleration parameter, and equation of state for this model are investigated.

\end{abstract}

\maketitle

\section{Introduction}

Einstein General Relativity (GR) is a successful theory, both theoretically and experimentally. GR is a classical theory that describes the gravitational interaction between all particles. It has been intensely tested since its construction until today \cite{Will}. However, there are many unanswered questions, such as, how a quantizable theory of gravity should be formulated? What is the nature of dark energy and dark matter? In addition, data from observational cosmology state that the universe has two phases of the cosmic acceleration. These phases are called inflation and late-time cosmic acceleration or dark energy. How to explain the two phases of the cosmic acceleration of the universe? As a consequence of the difficulties of solving these questions, alternative theories of GR have been developed. For a review see \cite{Harko, Albert, Gonzalo}. Here the $f(R)$ gravity theory is considered.

$f(R)$ gravity is known as the simplest modification of GR. In this theory the Einstein-Hilbert action is modified by replacing the Ricci scalar $R$ with an arbitrary function $f(R)$ \cite{Harko, Felice, Fara, Noj, Sot, Capoz}. The flexibility in defining the function $f(R)$ attracts a lot of interest since it allows the description of a wide range of cosmological phenomena \cite{Noj, Capoz}. For example, a natural modification is to add terms to the action like $f(R)\sim R^n$. For $n>1$, these terms lead to modifications of the standard cosmology at early times, specifically in the period known as inflation \cite{Star1, Vil}. For $n<1$, the corrections emerge as an important ingredient in explaining the late-cosmic acceleration of the universe \cite{Carroll1, Carroll2}. A well-consistent $f(R)$ model by combining models proposed by Starobinsky and Carroll–Duvvuri–Trodden–Turner (CDTT) model has been constructed \cite{Sharif, Kausar}. So this generalized model helps to cover all the expansion history of the universe. The main objective of this paper is to investigate some $f(R)$ models in the framework of a holographic dark energy model.

There are two different ways to explain the origin of the current acceleration of the Universe: (i) modified gravity theories and (ii) dark energy models. Dark energy is an unknown component that acts against the gravitational force, accelerating the expansion of the Universe. In order to solve the dark energy puzzle, the Holographic Dark Energy (HDE) hypothesis is a promising approach \cite{CKN, hoo, Suss}. Using the holographic principle, the standard holographic energy density depends upon the entropy-area relation of the black holes, i.e., $S\sim A$ with $A$ being the horizon area \cite{CKN, Li}. New HDE models can be proposed by using the holographic hypothesis and a generalized entropy. Gravity is a long-range interaction, then it can satisfy the non-extensive probability distributions. It is known that in non-additive systems, such as gravitational and cosmological, the usual Boltzmann-Gibbs additive entropy should be generalized to the non-extensive entropy. Here the non-extensive Tsallis entropy is considered \cite{Tsallis1, Tsallis2, Wilk, Tsallis3}.

From the holographic hypothesis and the Tsallis entropy, a new HDE model, called Tsallis Holographic Dark Energy (THDE), has been proposed \cite{Bamba2}. In this context, the horizon entropy of a black hole can be modified as $S=\gamma A^\delta$, where $\gamma$ is an unknown constant and $\delta$ is the non-additivity parameter. Note that, the Bekenstein entropy (i.e., $S\sim A$) is recovered at the appropriate limit. The Tsallis Holographic Dark Energy (THDE) density is $\rho\sim H^{4-2\delta}$, with $H$ being Hubble radius which plays the role of the IR cutoff. There are numerous applications that use THDE. For example, effects of considering various infrared (IR) cutoffs, including the particle horizon, Ricci horizon and Granda-Oliveros (GO) cutoffs have been explored \cite{Bamba}, cosmological consequences of THDE in the framework of Brans-Dicke gravity and modified Brans-Dicke gravity have been studied \cite{Valdir, Rani}, cosmological implications characterized by the energy density of THDE have been investigated \cite{Sar}, cosmic implications of THDE in a flat Friedmann–Robertson–Walker  universe in which there is no interaction between the cosmos sectors have been analyzed \cite{Bamba2}, a modified cosmological scenario that arises from the application of non-extensive thermodynamics with varying exponent has been presented \cite{Noj2}, THDE in the framework of dynamical Chern–Simon modified gravity and non-flat FRW universe has been discussed \cite{Ja}, some cosmological features of THDE in braneworld have been studied \cite{Valdir2},  $f(T)$ modified gravity model in the THDE framework has been considered \cite{AA}, $f(G,T)$ gravity and THDE has been analyzed \cite{Saba}, extended teleparallel gravity theory with Gauss–Bonnet term using holographic dark energy models has been reconstructed \cite{Jaw}, among others. In addition, this entropy is also confirmed in the framework of quantum gravity. For example, the consequences for the black hole area using a fractal structure for the horizon geometry has been investigated \cite{Barrow20}. The consequences and implications of this generalized entropy in cosmological setups have been studied \cite{Mora}. This shows that the generalized entropy may be in accordance with the thermodynamics laws, the Friedmann equation, and the universe expansion. In this paper, the THDE is considered and the evolution of the universe in the $f(R)$ gravity is obtained. The consequences of THDE on the Starobinsky and CDTT models are analyzed.

This paper is organized as follows. In section II, the $f(R)$ gravity is introduced. The field equations for a FRW space-time are obtained. Three different $f(R)$ models are discussed. In section III, a brief introduction to THDE is made. In section IV, the results are analyzed. The scale factor, the deceleration parameter, and the equation of state for different $f(R)$ models in the THDE context are investigated. In section V, some concluding remarks are presented.

\section{$f(R)$ gravity theory}

The modified gravity theory considered here is one of the simplest extensions of GR. In this theory, the Einstein-Hilbert lagrangian is generalized to be a general function $f$ of the Ricci scalar $R$, i.e.,
\bea
S = \frac{1}{2\kappa}\int f(R)\sqrt{-g}\,d^4x + S_M,
\eea
where $\kappa=8\pi G$, $G$ is the gravitational constant, $g$ is the determinant of the metric and $S_M$ is the action associated with the matter. In order to obtain the field equations of $f(R)$ gravity, the variation of the lagrangian with respect to the metric tensor is considered. Then
\bea
R_{\mu\nu}f'(R)-\frac{1}{2}g_{\mu\nu}f(R)-\nabla_\mu\nabla_\nu f'(R)+g_{\mu\nu}\Box f'(R)= \kappa\, T_{\mu\nu}, 
\eea
where a prime denotes differentiation with respect to $R$, $\nabla_\mu$ is the covariant derivative, $\Box\equiv\nabla_\mu\nabla^\mu $ and $T_{\mu\nu}$ is the energy-momentum tensor defined as
\bea
T_{\mu\nu}=-\frac{2}{\sqrt{-g}}\frac{\delta S_M}{\delta g^{\mu\nu}}. 
\eea
It is interesting to note that these equations are fourth-order differential equations in the metric. 

These equations in a form similar to the standard Einstein field equations are written as
\bea
G_{\mu\nu}=\kappa\mathbf{T}_{\mu\nu},\label{FE}
\eea
where $G_{\mu\nu}$ is the Einstein tensor and
\bea
\mathbf{T}_{\mu\nu}\equiv\frac{\kappa}{f'(R)}\left(T_{\mu\nu}+T_{\mu\nu}^{eff}\right),
\eea
with
\bea
T_{\mu\nu}^{eff}&=&\frac{1}{\kappa}\Bigl[\frac{1}{2}g_{\mu\nu}\left(f(R)-f'(R)R\right)+\nabla_\mu\nabla_\nu f'(R)-g_{\mu\nu}\Box f'(R)\Bigl],
\eea
being an effective energy-momentum tensor containing geometric terms. It does not satisfy any energy condition and its effective energy density is,
in general, not positive-definite.

To study the field equations of the $f(R)$ theory, eq. (\ref{FE}), the flat FRW universe is considered. The FRW space-time is described by the line element 
\bea
ds^2 = -dt^2 + a^2(t)\left[dr^2 + r^2d\theta^2 + r^2sin^2\theta d\phi^2\right],
\eea
where $a(t)$ is the scale factor. By taking the perfect fluid as the content of matter, which is given by energy-momentum tensor
\bea
T^{\mu\nu}=(\rho+p)u^\mu u^\nu + pg^{\mu\nu}
\eea
with $\rho$ and $p$ being the energy density and pressure of the fluid, respectively, the field equations become
\bea
H^2 &=& \frac{1}{3f'(R)}\left(\kappa\rho + \frac{Rf'(R)-f(R)}{2} - 3H\dot{R}f''(R)\right),\label{friedmann00}\\
2\dot{H}+3H^2 &=& -\frac{\kappa}{f'(R)}\Bigl(p - \frac{Rf'(R)-f(R)}{2} + 2H\dot{R}f''(R)+ \dot{R}^2f'''(R)+\ddot{H}f''(R)\Bigl),\label{friedmannii}
\eea
where $H=\frac{\dot{a}}{a}$ and an overdot denotes differentiation with respect to the time. These equations are known as the Friedmann equations. The eq. \eqref{friedmannii} can be written as
\bea
2\dot{H}+2H^2 &=& -H^2-\frac{\kappa}{f'(R)}\Bigl(p - \frac{Rf'(R)-f(R)}{2} + 2H\dot{R}f''(R)+ \dot{R}^2f'''(R)+\ddot{H}f''(R)\Bigl).
\eea
Using eq. \eqref{friedmann00} we get
\begin{equation}
\dot{H}+H^2 = \frac{\ddot{a}}{a} = -\frac{\kappa}{6}\left[\rho+\rho_{eff}+3\left(p+p_{eff}\right)\right],
\label{uniaofriedmann}
	\end{equation}
with
{\small
\begin{equation}
\rho_{eff} = \frac{1}{\kappa f'(R)}\left(\frac{Rf'(R)-f(R)}{2} - 3H\dot{R}f''(R)\right)
\end{equation}}
and
\bea
		p_{eff}= \frac{\kappa}{\kappa f'(R)}\Bigl(\dot{R}^2f'''(R)+\ddot{H}f''(R) - \frac{Rf'(R)-f(R)}{2} + 2H\dot{R}f''(R)\Bigl).
\eea

For an accelerated universe ($\ddot{a}>0$) the relation \eqref{uniaofriedmann} must satisfy
\begin{equation}
\rho+\rho_{eff}+3\left(p+p_{eff}\right)<0 \quad\implies\quad \frac{p+p_{eff}}{\rho+\rho_{eff}}<-\frac{1}{3}.
\end{equation}
Then an equation of state is constructed as
\begin{equation}
\omega_{tot}=\frac{p+\dot{R}^2f'''(R)+\ddot{H}f''(R) - \frac{Rf'(R)-f(R)}{2} + 2H\dot{R}f''(R)}{\rho+\frac{Rf'(R)-f(R)}{2} - 3H\dot{R}f''(R)},\label{parametrodeestado}
\end{equation}
where $\omega_{tot}\equiv\frac{p_{tot}}{\rho_{tot}}$ with $\rho_{tot}=\rho+\rho_{eff}$ and $p_{tot}=p+p_{eff}$.
It allows to interpret the behavior of the system from the $f(R)$ model and the matter content of the system.	

To investigate the evolution of the universe, given by the field equations, a particular model of $f(R)$ must be chosen. 

\subsection{$f(R)$ gravity models}

There are several $f(R)$ gravity models. Here models such as $f(R)\propto R^n$ are analyzed. For $n>1$, such terms lead to modifications of the
standard cosmology at early times which lead to de Sitter behavior \cite{Starobinsky1, Capo,Starobinsky2}. For $n<1$, such corrections become important in the late Universe and can lead to self-accelerating vacuum solutions \cite{Carroll1, Carroll2}. These models are briefly discussed.

\subsection{Starobinsky's Model}

In this case, the $f(R)$ function is provided as
\bea
f(R)\propto R^n
\eea
and a generic power law for the scale factor $a(t)\propto t^\alpha$ is considered. Then the effective equation of state parameter $\omega_{eff}$ becomes
\bea
\omega_{eff}=-\frac{6n^2-7n-1}{6n^2-9n+3},
\eea
with $n\neq 1$ and the $\alpha$ parameter is given in terms of $n$. Thus an appropriate choice of $n$ leads to the desired value of $\omega_{eff}$. For example, $n=2$ leads to $\omega_{eff}=-1$, that describes the Starobinsky inflation.

\subsection{Carroll-Duvvuri-Trodden-Turner (CDTT) Model}

In this model, the main idea is to show that the modification becomes important only in regions of extremely low space-time curvature. Here the $f(R)$ function is given as
\bea
f(R)=R-\frac{\mu^{2n+2}}{R^n},
\eea
where $\mu$ is a parameter with units of mass. By taking a power law for the scale factor, the $\omega_{eff}$ parameter becomes
\bea
\omega_{eff}=-1+\frac{2(n+2)}{3(2n+1)(n+1)}.
\eea
Then the case $n=1$ leads to $\omega_{eff}=-2/3$, which implies ultimate cosmic acceleration. 

\subsection{Generalized model}\label{GM}

This model is obtained by combining models proposed by Starobinsky and CDTT, also known as generalized CDTT (gCDTT) model \cite{Hu1, Hu2, Sharif}. Here the $f(R)$ function is defined as
\bea
f(R)=R+\lambda R^2-\sigma\frac{\mu}{R},\label{eq21}
\eea
where $\sigma=\pm 1$. The $\sigma=+1$ model leads to the generalized model that is composed of the Starobinsky and original CDTT models. However the $\sigma=-1$ model does not change the asymptotic behavior of the model ($\omega_{eff}=-2/3$), but it carries instabilities problems. Nevertheless, the CDTT model with $\sigma=-1$ is illustrative and its pathologies may be avoidable \cite{Hu1}.

Our main objective is to study the evolution of different $f(R)$ models, which describe different phases of the universe, in the presence of dark energy based on the holographic principle. In the next section, the holographic dark energy model is discussed.

\section{Tsallis holographic dark energy}

Here a brief introduction to the Holographic Dark Energy (HDE) model is presented. It is a promising candidate to resolve the dark energy puzzle. The holographic principle states that the number of degrees of freedom of a physical system should scale with its bounding area rather than with its volume \cite{hoo, Suss}. The holographic energy density is given as
\bea
\rho=\frac{3c^2M_p^2}{L^2}
\eea
and it depends on the entropy–area relationship of the black hole, i.e., $S\propto A$, where $A$ is the area of the event horizon of the black hole \cite{CKN}.

From the relation between the UV and IR cutoffs a new expression for the entropy is proposed \cite{CKN}
\bea
L^3\Lambda^3\leq S^{3/4},\label{Cohen}
\eea
where $L$ and $\Lambda$ are IR and UV-cutoffs, respectively. Here, the standard Boltzmann-Gibbs additive entropy is the one that it is
usually used. However, in systems including long-range interactions, like gravitational systems, one should probably use non-extensive statistics to study the systems. Then the Boltzmann-Gibbs additive entropy should be generalized to the non-extensive Tsallis entropy \cite{Tsallis1, Tsallis2, Wilk}, which can be applied in all cases, possessing the former as a limit. Tsallis and Cirto \cite{Tsallis3} suggested that the HDE can be redefined
\bea
S=\gamma A^\delta,\label{Tsallis}
\eea
where $\gamma$ is an unknown constant and $\delta$ denotes the non-additivity parameter. In the limit $\gamma=1/4G$ and $\delta=1$ the Bekenstein-Hawking entropy is recovered.

By combining eqs. (\ref{Cohen}) and (\ref{Tsallis}), the vacuum energy density is given as
\bea
\rho=B L^{2\delta-4}
\eea
with $B$ being an unknown parameter. It is Tsallis Holographic Dark Energy Density (THDE). Different IR cutoffs can be proposed to represent the accelerated expansion of universe, such as Hubble horizon, event horizon, particle horizon and so forth. By choosing the simplest IR-cutoff as Hubble horizon $(L=H^{-1})$, the energy density becomes
\bea
\rho=B H^{4-2\delta}.\label{Tis}
\eea
The $\delta$ parameter is related to the dimensionality $d$ of the system. It is defined as $\delta=\frac{d}{d-1}$ for $d>1$ \cite{Tsallis3}. It is important to note that, for $\delta=1$ the usual holographic dark energy is recovered. In addition, it is worth mentioning that in the special case $\delta=2$ the above relation gives the standard cosmological constant case, that is, $\rho=const.=\Lambda$.

In order to study the Friedman equation and the equation of state, determined in the previous section, the energy conservation,
\bea
\dot{\rho}+3H(\rho+p)=0,
\eea 
is used to obtain an expression for the pressure. Then
\bea
p = \frac{2\delta-4}{3}B\dot{H}H^{-2\delta+2} - BH^{-2\delta+4}.
\eea
In the next section, energy density and pressure are used in the Friedmann equation \eqref{friedmann00} and in the equation of state \eqref{parametrodeestado}.

\section{Results - THDE in $f(R)$ gravity}

The Friedmann equation and the equation of state, eq. (\ref{uniaofriedmann}) and eq. (\ref{parametrodeestado}) respectively, are complicated. Then due to the complexity of these equations, a simple, closed-form solution cannot be obtained by purely analytical means. Given this difficulty, these equations are solved numerically. In addition, the deceleration parameter $q(t)$ is also calculated. The deceleration parameter is defined as
\bea
q(t)\equiv-\frac{\ddot{a}a}{\dot{a}^2}.
\eea

By taking the universe with a holographic dark energy density, as given by eq. (\ref{Tis}), the main objective is to analyze the equations of the generalized models of $f(R)$, eq. (\ref{eq21}). 

First, let's investigate the two models ($\sigma=\pm 1$), for an empty universe, that is, THDE energy density is zero,  $\rho=0$. The Fig. 1 presents the behavior of the parameter $\omega_{eff}$ for these models.
\begin{figure}[h]
\includegraphics[scale=0.3]{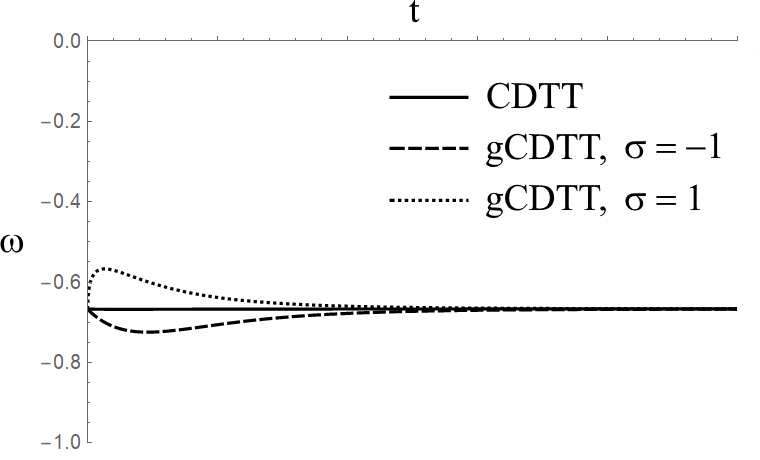}
\caption{ $\omega_{eff}$ parameter for the CDTT and gCDTT models. This result is for a particular choice of $a(0)$, $\dot{a}(0)$, $\ddot{a}(0)$ and a specific value of $\lambda=1,05\times10^7$ and $\mu=4\times10^{-9}$.}
\end{figure}

In the CDTT model, regardless of the defined signal, the behavior is a quadratic expansion ($\omega=-2/3$). Adding the Starobinsky term, a positive or negative contribution to the acceleration appears.

Now, the generalized CDTT model (for both cases $\sigma=\pm 1$ ) under the influence of THDE, eq. (\ref{Tis}), is analyzed. Different values of $\delta$, such as $\delta=1$, for which the Bekenstein-Hawking entropy is recovered,  $\delta_{d=3}=3/2$, $\delta_{d=4}=4/3$, for the dimensions 3 and 4 respectively, are considered. Using the THDE energy density and pressure, the Friedmann equation becomes
\bea
H^2 = \frac{1}{3f'}\left(\kappa BH^{-2\delta+4} + \frac{Rf'-f}{2} - 3H\dot{R}f''\right)
\eea
and the equation of state is 
\bea
\omega = \frac{\frac{2\delta-4}{3}B\dot{H}H^{-2\delta+2} - BH^{-2\delta+4}+\dot{R}^2f'''+\ddot{H}f'' - \frac{Rf'-f}{2} + 2H\dot{R}f''}{BH^{-2\delta+4}+\frac{Rf'-f}{2} - 3H\dot{R}f''} .
\eea
In addition, the deceleration parameter $q(t)$ is also calculated.

\subsection{$f(R)=R+\lambda R^2+\mu/R$}

First, the evolution of the scale factor in this gravitational model with THDE is obtained. Four different cases are compared, $\delta=1$, $\delta_{d=3}=3/2$, $\delta_{d=4}=4/3$ and $\rho=0$, i.e., THDE is absent. In Fig. 2 is shown that THDE positively affects the expansion rate of the universe.
\begin{figure}[h]
\includegraphics[scale=0.3]{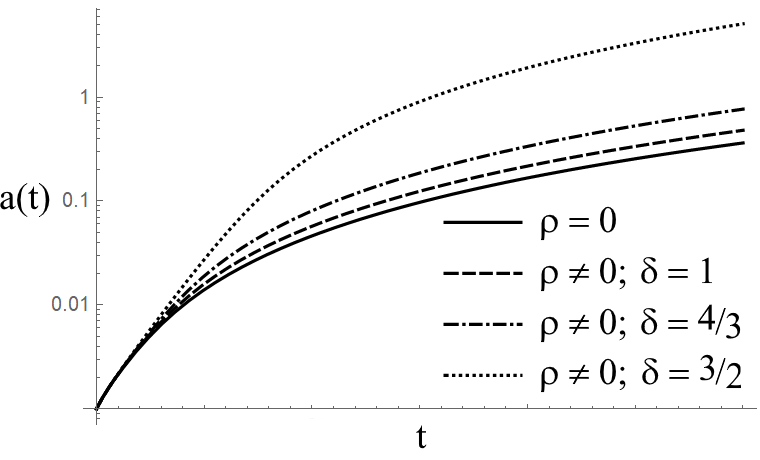}
\caption{ Scale factor for different energy densities and different cases of $\delta$ in gCDTT model with $\sigma=-1$. }
\end{figure}

Now, the equation of state and the deceleration parameter are calculated. These parameters are displayed in Fig. 3 and Fig. 4, respectively.
\begin{figure}[h]
		\begin{minipage}{0.45\textwidth}
			\centering
			\includegraphics[scale=0.3]{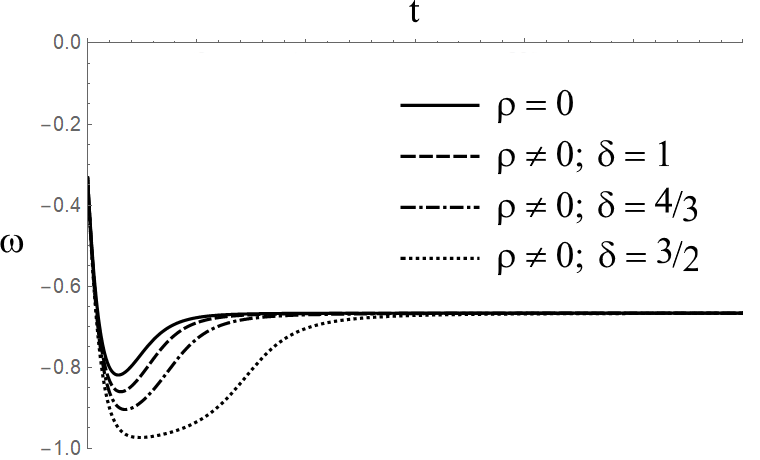}
			\caption{$\omega_{eff}(t)$ parameter for different energy densities and different cases of $\delta$.} \label{omega}
		\end{minipage}
		\quad\quad\quad\quad
		\begin{minipage}{0.45\textwidth}
			\centering
		\includegraphics[scale=0.3]{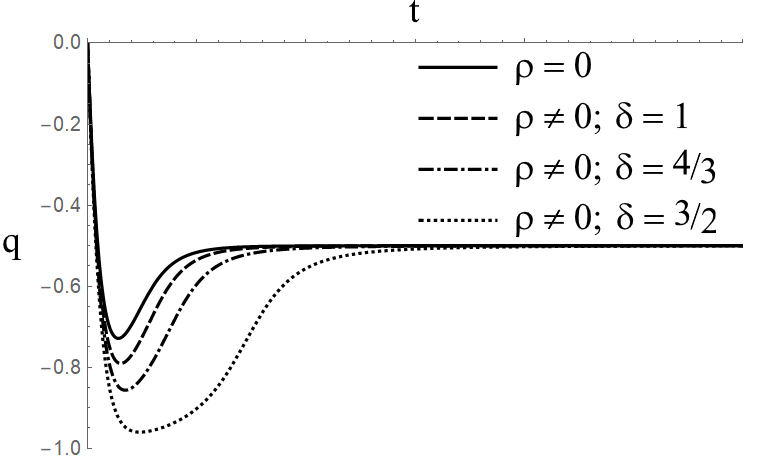}
		\caption{$q(t)$ - Deceleration parameter for different energy densities and different cases of $\delta$.} \label{q(t)}
		\end{minipage}
	\end{figure}
Our results exhibit that the THDE intensifies the initial acceleration. Note that, in the purely effective case ($\rho=0$) the system initially expand as a de Sitter expansion ($\omega_{eff}=-1$) and then as a power law ($\omega_{eff}=-2/3$). This result is expected. However, in the cases with THDE ($\rho\neq 0$) the de Sitter behavior is modified. Therefore, THDE energy density contributes positively to the initial exponential expansion of the universe. In addition, the THDE does not contribute for the late accelerated expansion of the universe. Furthermore, the $q(t)$ results confirm the results obtained for the $\omega_{eff}$.
	
\subsection{$f(R)=R+\lambda R^2-\mu/R$}	

Here, the same parameters $a(t)$, $\omega_{eff}$ and $q(t)$ are analyzed. The scale factor is shown in the Fig. 5.
\begin{figure}[h]
\includegraphics[scale=0.3]{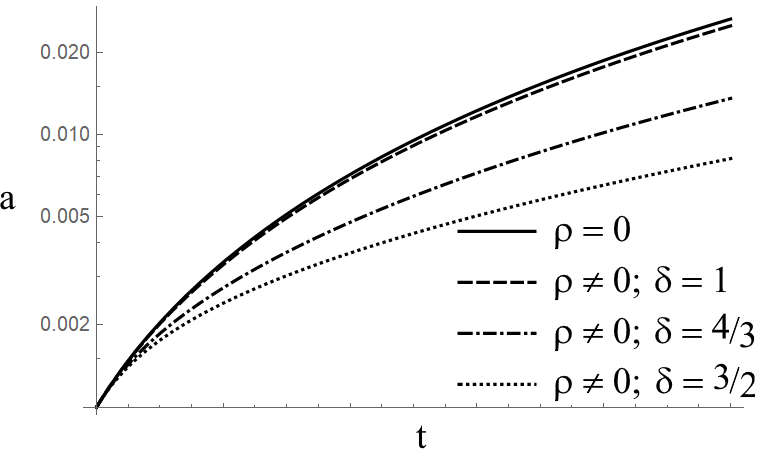}
\caption{ Scale factor for different energy densities and different cases of $\delta$ in gCDTT model with $\sigma=+1$. }
\end{figure}
In this scenario, THDE energy density contributes to decreasing the rate of expansion. 

The parameters $\omega_{eff}$ and $q(t)$ are presented in Fig. 6 and Fig. 7, respectively.
\begin{figure}[h]
		\begin{minipage}{0.45\textwidth}
			\centering
			\includegraphics[scale=0.3]{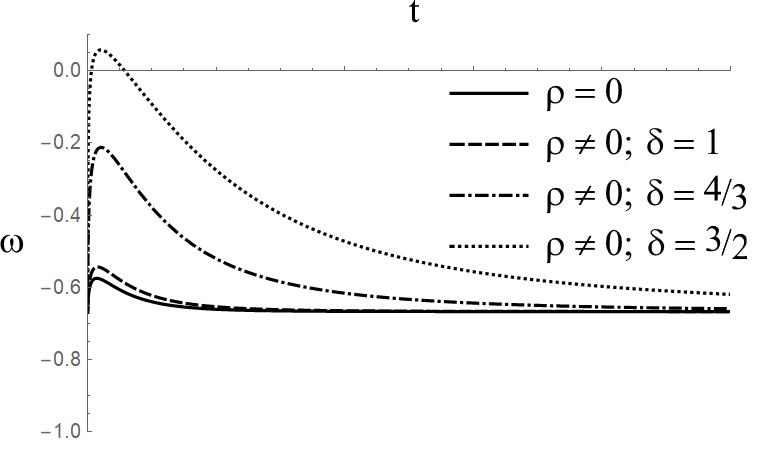}
			\caption{$\omega_{eff}(t)$ parameter for different energy densities and different cases of $\delta$.} 
		\end{minipage}
		\quad\quad\quad\quad
		\begin{minipage}{0.45\textwidth}
			\centering
		\includegraphics[scale=0.3]{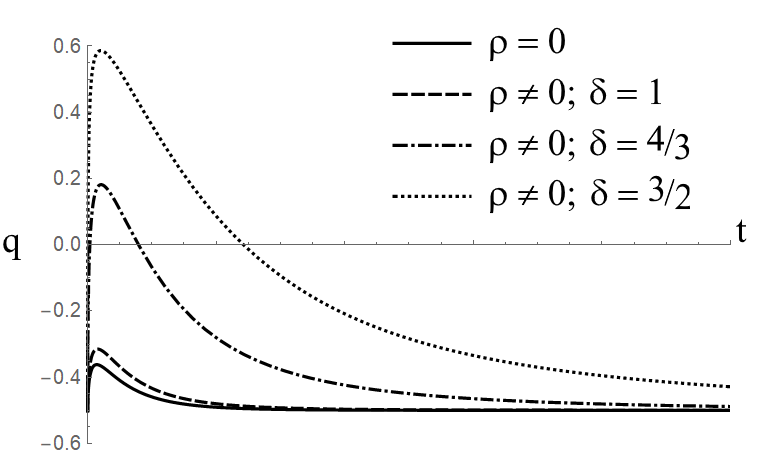}
		\caption{$q(t)$ - Deceleration parameter for different energy densities and different cases of $\delta$.} 
		\end{minipage}
	\end{figure}
A slowdown occurs at the beginning of the evolution of the system. This deceleration is a characteristic of the $f(R)$ gravity model. Although there is a change in the initial expansion, the asymptotic behavior remains a power law.

Therefore, in both generalized models $\sigma=\pm 1$, the THDE changes only the initial expansion that is described by the Starobinsky's model.

In order to obtain a complete analysis, our main results are also plotted in terms of the redshift $z$, which is defined as $1+z=\frac{a(t_0)}{a(t)}$. The equation of state $\omega_{eff}(z)$ and the deceleration parameter $q(z)$ for the model $f(R)=R+\lambda R^2+\mu/R$ are shown in Figs. 8 and 9.  While Figs. 10 and 11 display the behavior of these parameters for the model $f(R)=R+\lambda R^2-\mu/R$. It is important to note that, the physical interpretation for these results is the same as discussed for Figures 3, 4, 6 and 7.
\begin{figure}[h]
		\begin{minipage}{0.45\textwidth}
			\centering
			\includegraphics[scale=0.3]{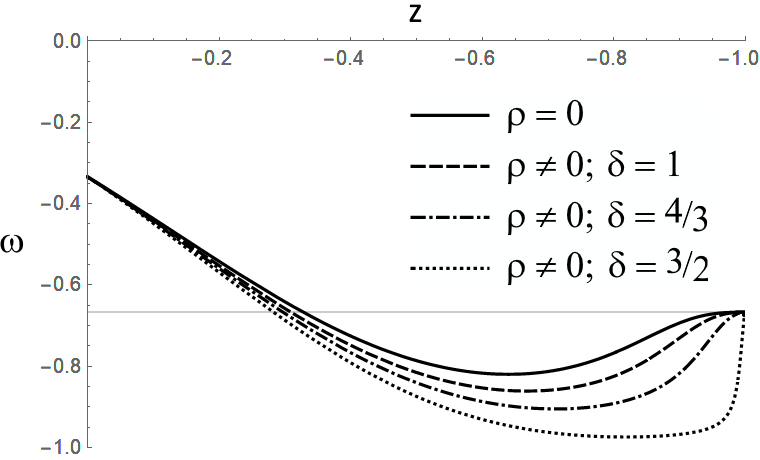}
			\caption{$\omega_{eff}(z)$ parameter for the model $f(R)=R+\lambda R^2+\mu/R$.} 
		\end{minipage}
		\quad\quad\quad\quad
		\begin{minipage}{0.45\textwidth}
			\centering
		\includegraphics[scale=0.3]{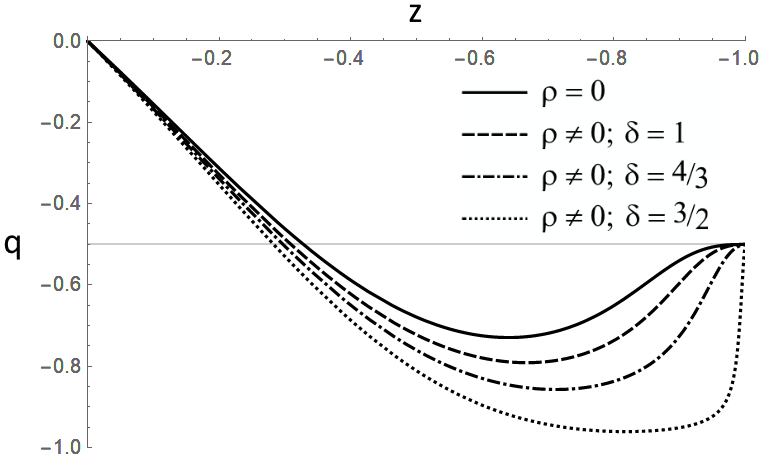}
		\caption{$q(z)$ - Deceleration parameter for the model $f(R)=R+\lambda R^2+\mu/R$.} 
		\end{minipage}
	\end{figure}

\begin{figure}[h]
		\begin{minipage}{0.45\textwidth}
			\centering
			\includegraphics[scale=0.3]{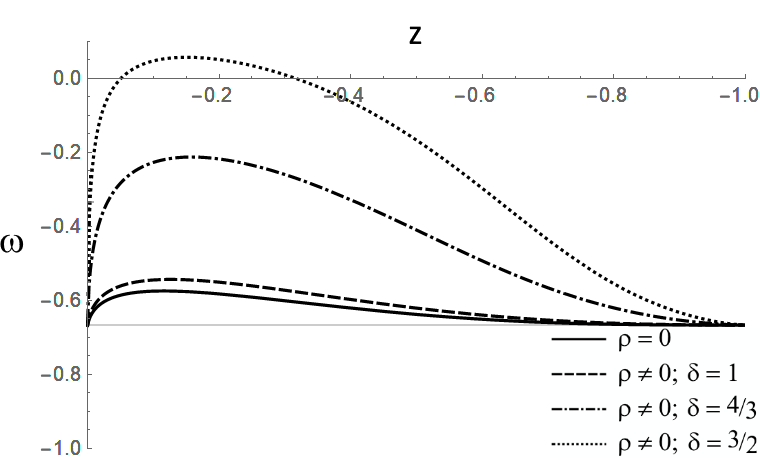}
			\caption{$\omega_{eff}(z)$ parameter for the model $f(R)=R+\lambda R^2-\mu/R$.} 
		\end{minipage}
		\quad\quad\quad\quad
		\begin{minipage}{0.45\textwidth}
			\centering
		\includegraphics[scale=0.3]{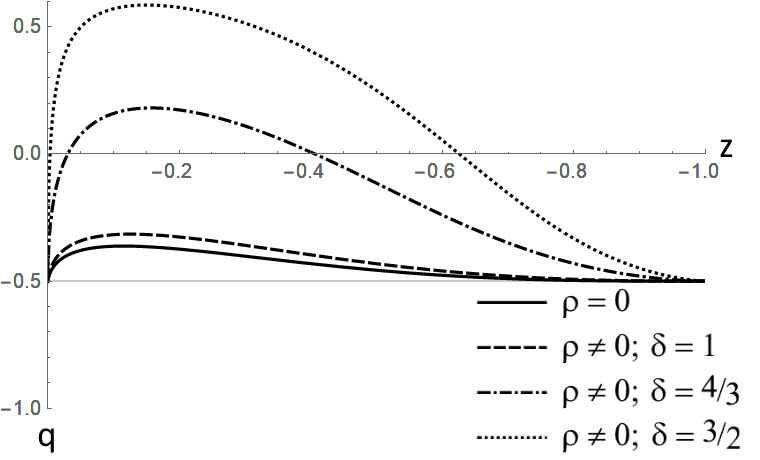}
		\caption{$q(z)$ - Deceleration parameter for the model $f(R)=R+\lambda R^2-\mu/R$.} 
		\end{minipage}
	\end{figure}

It is interesting to observe that, our results are obtained for the cases: (a) $\rho=0$, i.e. an empty universe and (b) $\rho\neq 0$, with $\rho$ being the Tsallis Holographic Dark Energy density. In order to extend our study, let us consider adding a pressureless fluid to the matter content. As an example, the equation of state $\omega_{eff}(z)$ for both models, i.e. $\sigma=\pm$, are shown in Figs. 12 and 13. In this analysis, the following cases are considered: (i) $\rho=0$, an empty universe; (ii) $\rho=\rho_M$, a universe filled with a pressureless fluid; (iii) $\rho=\rho_M+\rho_{THDE}$, a universe filled by THDE and pressureless fluid, simultaneously and (iii) $\rho=\rho_{THDE}$, a universe filled only with THDE. Here, as an example, the case $\delta=3/2$ has been considered. Our results display that the presence of pressureless fluid changes the initial acceleration of the universe. This is an expected result.

\begin{figure}[h]
		\begin{minipage}{0.45\textwidth}
			\centering
			\includegraphics[scale=0.3]{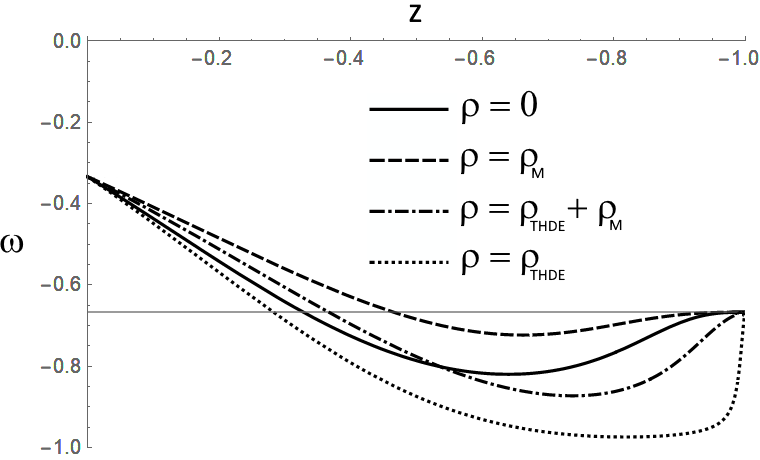}
			\caption{$\omega_{eff}(z)$ parameter for the model $f(R)=R+\lambda R^2+\mu/R$. Here, a pressureless fluid has been also considered as matter content.} 
		\end{minipage}
		\quad\quad\quad\quad
		\begin{minipage}{0.45\textwidth}
			\centering
		\includegraphics[scale=0.3]{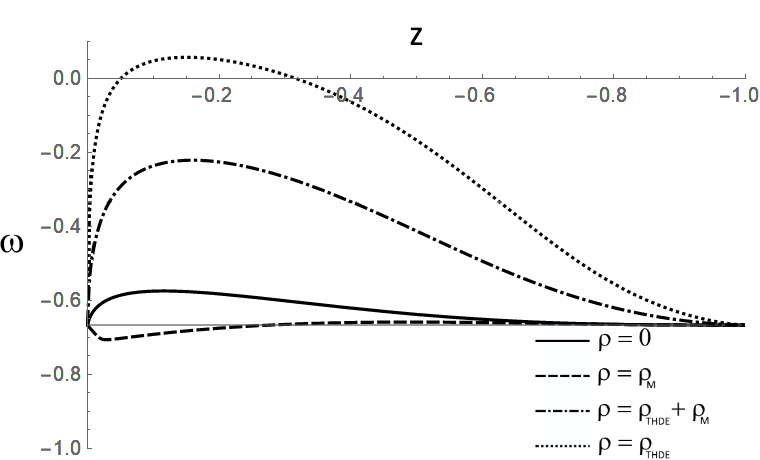}
		\caption{$\omega_{eff}(z)$ parameter for the model $f(R)=R+\lambda R^2-\mu/R$. Here, a pressureless fluid has been also considered as matter content.} 
		\end{minipage}
	\end{figure}

\section{Conclusions}

The most well-known and tested gravitational theory is GR. However, it is not a fundamental theory and there are problems that it does not explain. This leads to modified theories of gravity. Here the $f(R)$ theories of gravity are considered. In this context, a holographic dark energy model is studied. Using the holographic principle and Tsallis entropy, the Tsallis Holographic Dark Energy (THDE) is constructed. Here, Friedmann equation and the equation of state, for different $f(R)$ theories, in a universe with THDE is solved. Our main results are obtained for the combined model $f(R)=R+\lambda R^2-\sigma\frac{\mu}{R}$, where $\sigma=\pm 1$. It consists of the Starobinsky and CDTT models. Then it describes the entire history of the expansion of the universe. The evolution of the scale factor in the presence of THDE is changed for both models $\sigma=\pm 1$. For the deceleration parameter and for the $\omega_{eff}$ parameter the result is very interesting. It is shown that THDE changes the behavior in the first phase of the acceleration called inflation, described by Starobinsky's model. While for the late-cosmic acceleration, described by the CDTT model, is not modified due to THDE. Therefore, THDE can change the $f(R)$ gravity behavior, more specifically in the standard cosmology at early times. In addition, this modification depends on the non-additivity parameter $\delta$. Furthermore, it is interesting to note that the phase transition between the two models is altered when THDE is present.

\acknowledgments
This work by A. F. S. is supported by CNPq projects 308611/2017-9 and 430194/2018-8; P. S. E. thanks CAPES for financial support.

\end{document}